\documentclass[aps,twocolumn,showpacs]{revtex4}

\usepackage{graphicx}
\usepackage{amssymb}
\usepackage{amsmath}

\begin{document}

\title{Surface criticality in random field magnets}

\author{L. Laurson and  M.\ J.\ Alava}

\affiliation{Helsinki University of Technology, Laboratory of Physics,\\ 
P.O.Box 1100, FIN-02015 HUT, Finland }

\begin{abstract}
\noindent
The boundary-induced scaling of three-dimensional 
random field Ising magnets is investigated close to the bulk
critical point by exact combinatorial optimization methods. 
We measure several exponents describing surface criticality: $\beta_1$ 
for the surface layer magnetization and the surface excess exponents 
for the magnetization and the specific heat, $\beta_s$  and $\alpha_s$. 
The latter ones are related to the bulk phase transition by the same scaling 
laws as in pure systems, but only with the same violation of 
hyperscaling exponent $\theta$ as in the bulk. The boundary 
disorders faster than the bulk, and the experimental and theoretical
implications  are discussed. 
\end{abstract}
\maketitle

\date{\today}

\section{Introduction}

The presence of quenched randomness leads to many differences in the 
statistical behavior if compared to ``pure systems''. 
This is true in many phenomena as transport properties in, 
for instance, superconductors, or 
in a rather wide range of cases in magnetism. 
Consider a domain wall in a magnet,
which gets pinned due to impurities. The scenario may vary according
to the symmetries of the system and to the character of the disorder,
but is described, in most general terms, by an
``energy landscape'' which develops a rich structure due to the
the presence of pinning defects \cite{generic}. 

The most usual and convenient example of such magnets
is given by the Ising model 
-universality class. Disorder is normally introduced as frozen
random bond'' and ``random field'' impurities, which can
change dramatically the nature of the phases of the model
and the character of the phase transition. Strong
enough bond disorder creates a spin glass -state, while the
random fields couple directly to the order parameter, the 
magnetization. 

The criticality in such models is usually studied by finite size scaling,
to extract the thermodynamic behavior. However,
real (experimental) systems are finite and have boundaries.
These break the translational invariance and create differences
in the critical behavior between the boundary region and the bulk.
The related phenomenon is called ``surface criticality'', and essential 
is that a whole set of new critical exponents arises, to describe the
behavior of various  quantities at and close to
surfaces \cite{ptcp8,ptcp10}. Here,
we investigate by scaling arguments and exact numerical methods
this phenomenon in the case of the random field Ising model (RFIM),
in three dimensions (3d). In this case, the RFIM has a bulk phase 
transition separating ferromagnetic and paramagnetic states.

The central question that we want to tackle is: how do
disorder and the presence of boundaries combine, in a system where
the critical bulk properties are already different from pure systems?
Though disordered magnets have been investigated earlier for the case of
weak bond-disorder \cite{selke,pleimling}, both spin-glasses - a 
possible future extension of our work
 - and the RFIM have not been studied \cite{heiko}. 
One general problem of the 3d RFIM has been how to observe
the critical behavior, and understanding the
boundary critical behavior provides an independent, novel
avenue for such purposes \cite{belanger,antifm,kleemann}.   
Such experiments are done on a number of systems from diluted antiferromagnets
in a field, \cite{belanger,antifm}, to binary liquids in porous
media, \cite{dierker}, and to relaxor ferroelectrics \cite{kleemann}. 

The particular characteristics of the RFIM is a 
complicated energy landscape, which manifests itself e.g.
in the violation of the usual hyperscaling relation of thermodynamics, 
and in the existence of an associated violation exponent $\theta$
and several consequences thereof. 
This is analogous to, 
for instance, spin glasses, and furthermore for 
surface criticality  presents the question how the broken translational
invariance combines with the energy scaling. Our results imply that
this can be understood by scalings that include
both the  bulk correlation length exponent $\nu$ and the
bulk $\theta$ and novel surface exponents.
Moreover, though the bulk RFIM 3d phase transition has been notoriously
difficult experimentally, the boundary order
parameter, say, should be quite sensitive to the control
one (temperature, in experiments and disorder here) 
and promises thus to make the surface criticality 
experimentally observable.

In the next section we overview the theoretical picture,
as applied to the RFIM. Section 3 presents the numerical
results, where the emphasis is two-fold. We discuss the
surface criticality on one hand, and on the other hand 
the decay of a surface field induced perturbation is analyzed,
since it has characteristics peculiar to a disordered magnet,
in contrast to pure systems. Finally, Section 4 finishes the
paper with a discussion of the results and future prospects.

\section{Surface criticality}

The RFIM Hamiltonian with a free surface $S$ reads
\begin{equation}
\label{RFIM_Ham}
H_{RFIM} = -J\sum_{\langle i,j\rangle \notin S}\sigma_i \sigma_j
-J_1\sum_{\langle i,j\rangle \in S}\sigma_i \sigma_j
- \sum_{i}h_i\sigma_i,
\end{equation}
where $J$ is the bulk (nearest neighbour) interaction strength while 
$J_1$ describes the strength of the {\em surface interaction}, in 
general different from $J$. $\sigma_i$ take the
values $\pm 1$. For simplicity, the random fields $h_i$ obey a 
Gaussian probability distribution $
P(h_i) = \frac{1}{\sqrt{2\pi}\Delta}
\exp{\left[-\frac{1}{2}\left(\frac{h_i}{\Delta}\right)^2\right]}$,
with a zero mean and standard deviation $\Delta$. One might have 
also external fields such as a bulk magnetic field $h$ and a surface 
magnetic field $h_1$ at $S$.

Being governed by a zero temperature fixed point, the phase transition
of the 3d RFIM can also be studied at $T=0$, where it takes place
at a critical $\Delta_c$. The transition is
of second order though it also exhibits some first-order characteristics:
the order parameter exponent $\beta$ is very close to zero 
\cite{middleton,rieger,hartmann_m}. The surface criticality of
the 3d RFIM is simplified by the fact that the lower critical
dimension is two \cite{aizenman,uusi}, thus in the absence of a surface 
magnetic field $h_1$ just an {\em ordinary transition} can take place. 
The surface orders only because the bulk does so, and the transition 
point is the bulk critical point.

Even in this case, there is a wide variaty of surface quantities.
Derivatives of the {\em surface free energy} $f_s$ (surface ground state
energy at $T=0$) with respect to surface fields, as the surface magnetic 
field $h_1$, yield {\em local quantities} (e.g. the surface layer 
magnetization $m_1=-\partial f_s/\partial h_1$), while 
derivatives of $f_s$ with respect to
bulk fields produce {\em excess quantities}, such as the excess magnetization
$m_s = -\partial f_s/\partial h$, defined by
\begin{equation}
\frac{1}{V}\int d^dx \ m({\bf x}) = m_b + \frac{S}{V}m_s + O(L^{-2}),
\end{equation}
where $m(\bf x)$ is the (coarse grained) magnetization at ${\bf x}$ and 
$V \sim L^d$ and $S$ are the sample volume and its surface area,
respectively. One also obtains
{\em mixed quantities} by taking second or higher derivatives of $f_s$.
We focus on the critical behavior of the local and the excess magnetization 
($m_1$ and $m_s$) as well as the excess specific heat $C_s$.

The RFIM bulk critical exponents are related via the
usual thermodynamic scaling relations, see Table \ref{table1}.
The hyperscaling relations, 
however, have the modified form
\begin{equation}
\label{hyper}
2-\alpha = \nu(d-\theta),
\end{equation}
with the additional exponent $\theta$ \cite{braymoore}. The usual way to 
relate the surface excess exponents to bulk exponents is to note that from 
the conventional hyperscaling (Eq. (\ref{hyper}) with $\theta = 0$) it 
follows that the singular part of the bulk free energy $f_b^{(sing)}$ scales 
with the correlation length $\xi$ as $f_b^{(sing)} \sim \xi^{-d}$. By 
making the analogous assumption for the surface free energy, $f_s^{(sing)} 
\sim \xi^{-(d-1)}$, one finds \cite{ptcp10}
\begin{equation}
\label{pure_scaling}
\alpha_s = \alpha + \nu, \quad \beta_s = \beta - \nu.
\end{equation}
In the case of the RFIM the above becomes less clear: does
the $\theta$-exponent get modifed? We 
assume that the exponent $\theta'$ in 
$f_s^{(sing)} \sim \xi^{-(d-1-\theta')}$ may in general be different from
the bulk exponent $\theta$, and obtain
\begin{eqnarray}
\label{alpha_s}
\alpha_s &=& \alpha + \nu - \nu(\theta-\theta'), \\
\label{beta_s} 
\beta_s &=& \beta - \nu + \nu(\theta-\theta').
\end{eqnarray}
To derive Eq. (\ref{beta_s}), the scaling form
$\frac{E_s^{(sing)}}{J} \sim t^{2-\alpha_s}
\tilde{E}_s [h/Jt^{-(\gamma+\beta)}]$ 
is used for the singular part of the excess ground state energy density 
$E_s^{(sing)}$
(which takes the role of the excess free energy at $T=0$), with 
$t \equiv (\Delta-\Delta_c)/J$, Eq. (\ref{alpha_s}) and the Rushbrooke
scaling law $\alpha + 2\beta + \gamma = 2$.
$\gamma$ is the exponent describing the critical 
behavior of the bulk susceptibility. 
Scaling relations relating $\beta_1$ to other 'local' surface exponents
can also be derived, but it cannot be expressed in terms of bulk exponents
alone.

\begin{table}[fht]
\begin{tabular}{lll}
\hline
Quantity & Definition & Exponent\\
\hline




excess magnetization & $m_s=-\frac{\partial f_s}{\partial h}$ & 
$m_s \sim (-t)^{\beta_s}$ \\

excess specific heat & 
$C_s=\frac{\partial^2 f_s}{\partial \Delta \partial J}$ & 
$C_s \sim |t|^{-\alpha_s}$ \\


surface magnetization & $m_1=-\frac{\partial f_s}{\partial h_1}$ & 
$m_1 \sim (-t)^{\beta_1}$ \\


\hline
\end{tabular}
\caption{Surface quantities
in terms of the surface free energy $f_s$, and the corresponding 
critical exponents ($t \equiv (\Delta-\Delta_c)/J$). Note that $T=0$
so that one uses instead of a free energy the
ground state energy.}
\label{table1}
\end{table}

\section{Numerical results}

\begin{figure}[ht]
\includegraphics[width=7cm
]{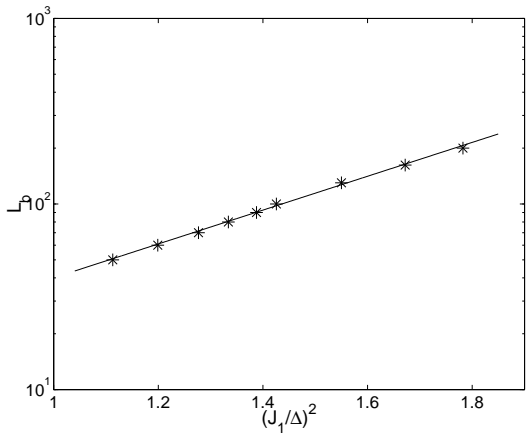}
\caption{The break-up length scale $L_b$ of the 2d surface 
layer of the 3d RFIM with a strongly paramagnetic bulk, $J=0.05 \Delta$, 
vs $(J_1/\Delta)^2$. $L_b$ is estimated by looking for a value of $J_1$
such that the surface will be totally ordered with probability $1/2$ 
while keeping $\Delta$ and $L$ fixed. The solid line corresponds to $A=2.1$.}
\label{lb}
\end{figure}

The exact ground state (GS) calculations are based on the 
equivalence of the $T=0$ RFIM with the maximum flow problem
in a graph \cite{alavaptcp}; we use a polynomial 
push-relabel preflow-type algorithm \cite{goldberg,seppala_vk}. 
If not stated otherwise, we study cubic systems of size $L^3$,  $L\leq100$. 
Free boundary conditions are used in one direction 
(the free surface under study) while in the remaining ones
periodic boundary conditions are imposed. The maximal
statistical error in what follows is of the order of the symbol size used,
so the error bars are omitted. Note that since in the present case
only the ordinary transition is possible,  
the critical exponents should be independent of the surface interaction $J_1$.
Complications arise, however, since in 2d the RFIM is effectively ferromagnetic
below the break-up length scale $L_b$, which scales as 
$L_b \sim \exp{[A(J/\Delta)^2]}$  
(see Fig. \ref{lb}) \cite{seppala_Lb, binder}.
This means that the surfaces have a tendency to be ordered ``an sich'',
and to see the true ordinary transition behavior, one needs $L > L_b$.
Thus, we use substantially weakened surface interactions $J_1 \ll J$
to circumvent this problem.

\begin{figure}[ht]
\includegraphics[width=7cm
]{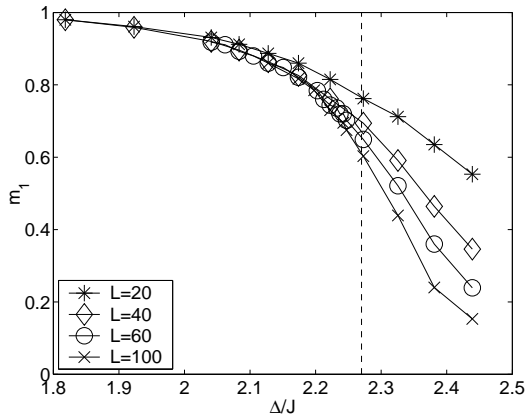}
\caption{Mean absolute value of the surface layer magnetization $m_1$ 
as a function of $\Delta/J$ for various $L$, $J_1=J$. The dashed vertical 
line corresponds to the critical point of the infinite system, 
$\Delta/J=2.27$.}
\label{m1}
\end{figure}

\subsection{Surface layer magnetization}

\begin{figure}[ht]
\includegraphics[width=7cm
]{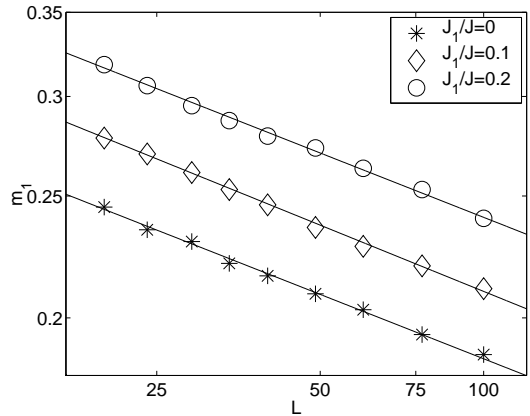}
\caption{A log-log plot of the surface layer magnetization $m_1$ as a function
of the system size $L$ at criticality, $\Delta/J = 2.27$, for various 
$J_1/J \ll 1$. The solid lines depict fits, with
$\beta_1/\nu = 0.17 \pm 0.01$ for all three cases shown}
\label{m1_crit}
\end{figure}

\begin{figure}[ht]
\includegraphics[width=7cm
]{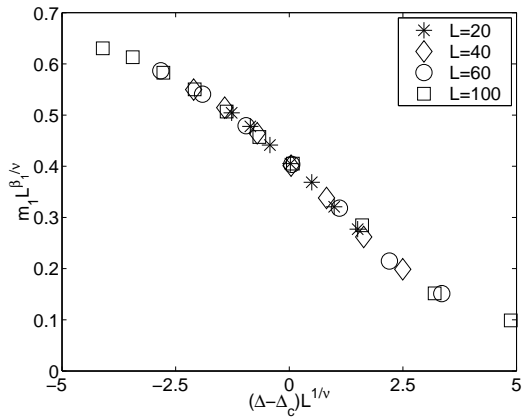}
\caption{A scaling plot of the surface layer magnetization $m_1$ in the 
case $J_1=0$, $J=1$, using $\Delta_c=2.27$, $\nu =1.37$ and $\beta_1=0.23$.}
\label{collapse}
\end{figure}

Fig. \ref{m1} shows an example of the magnetization $m_1$ of the surface layer
close to $\Delta_c$, obtained directly from the spin structure of the GS. We 
assume the finite size scaling ansatz
\begin{equation}
\label{m1_ansatz}
m_1 = L^{-\beta_1/\nu} \tilde{m}_1[(\Delta-\Delta_c)L^{1/\nu}],
\end{equation} 
where $\tilde{m}_1$ is a scaling function. At the critical point 
$\Delta=\Delta_c$, Eq. (\ref{m1_ansatz}) reduces to 
$m_1 \sim L^{-\beta_1/\nu}$. Fig. \ref{m1_crit} is a double logarithmic 
plot of $m_1$ versus $L$ at $\Delta_c/J = 2.27$ for three $J_1$-values.
All three are consistent with
\begin{equation}
\beta_1/\nu = 0.17 \pm 0.01.
\end{equation}
Using the bulk value $\nu=1.37 \pm 0.09$ \cite{middleton}, one obtains
\begin{equation}
\beta_1 = 0.23 \pm 0.03.
\end{equation}
Fig. \ref{collapse} depicts
$m_1L^{\beta_1/\nu}$ versus $(\Delta-\Delta_c)L^{1/\nu}$,
and with $\beta_1/\nu=0.17$, $\nu=1.37$ and $\Delta_c/J = 2.27$
one indeed obtains a decent data collapse. With $J_1 \approx J$, 
however, plotting $m_1(\Delta_c)$ versus $L$ produces a slightly 
different exponent, $\beta_1/\nu \approx 0.15$, and we could not 
get good data collapses, probably due to the fact that $L_b$ is 
large.

\subsection{Surface excess magnetization}

\begin{figure}[ht]
\includegraphics[width=7cm
]{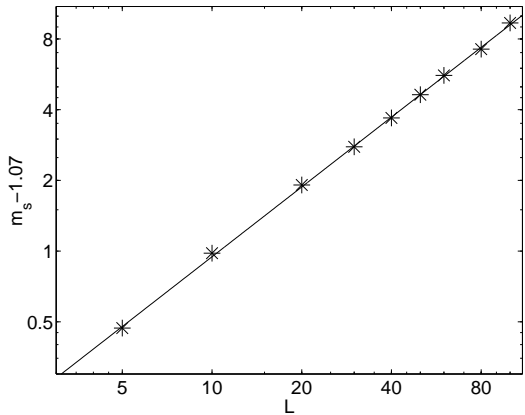}
\caption{A log-log plot of the excess magnetization $m_s$ as a function of 
the system size $L$ for $\Delta/J=2.27$, $J_1/J=0.1$. A background 
term of magnitude 1.07 has been substracted from $m_s$ to see the power-law 
behavior. The solid line is a power-law fit, with
$-\beta_s/\nu = 0.99$.}
\label{ms_crit}
\end{figure}

For the surface excess magnetization $m_s$, we use the finite size scaling
ansatz
\begin{equation}
m_s = L^{-\beta_s/\nu}\tilde{m}_s[(\Delta-\Delta_c)L^{1/\nu}],
\end{equation}
where $\tilde{m}_s$ is a scaling function. Since $\beta_1$ was found to be
independent of $J_1/J$ as long as $J_1/J \ll 1$ (in the limit $L\rightarrow 
\infty$, the independence of the exponents on $J_1/J$ should hold for
{\em any} $J_1/J$), one expects the same to apply for the
other exponents as well and we thus consider here only the case 
$J_1/J=0.1$. At the critical point, $m_s$ grows almost 
linearly with $L$ (Fig. \ref{ms_crit}), with
the exponent $-\beta_s/\nu = 0.99 \pm 0.02$. This 
yields, by again using $\nu = 1.37 \pm 0.09$,
\begin{equation}
\beta_s = -1.4 \pm 0.1.
\end{equation}

\subsection{Surface specific heat}

\begin{figure}[ht]
\includegraphics[width=7cm
]{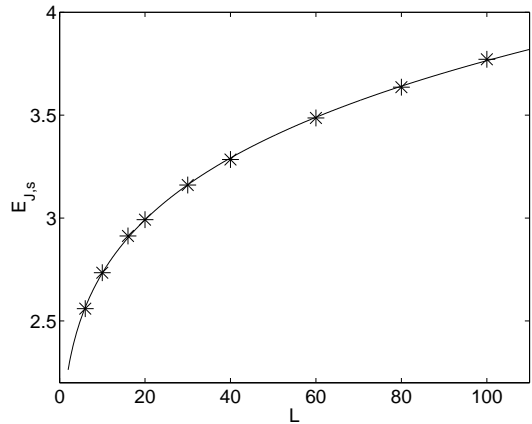}
\caption{A plot of the absolute value of the excess bond energy $E_{J,s}$ 
as a function of  $L$ for $\Delta/J=2.27$, $J_1/J=0.1$. 
The solid line corresponds to a fit of the form of Eq. (\ref{bondE_form}), with
$c_1=1.1292$, $c_2=0.9756$ and $(\alpha_s-1)/\nu=0.22.$}
\label{Cs_crit}
\end{figure}

In GS calculations, the specific heat is computed 
(recall $T=0$) by replacing the second derivative of
the free energy $f$ with respect to the temperature by 
the second derivative of the GS energy density $E$ 
with respect to $\Delta$ or $J$ \cite{hartmann}.
$\partial E / \partial J$ is the 
the bond part of $E$,
$E_J = L^{-d}\sum_{\langle i,j \rangle} \sigma_i \sigma_j$.
The excess specific heat exponent 
$\alpha_s$ is estimated according to Ref. \cite{middleton} (where 
the bulk one was considered). 
The singular part of the excess specific heat obeys
\begin{equation}
C_s^{(sing)} = L^{\alpha_s/\nu}\tilde{C}_s[(\Delta-\Delta_c)L^{1/\nu}],
\end{equation}
from which by integration it follows for the
singular part of the excess bond energy at criticality,
\begin{equation}
\label{bondE_form}
E_{J,s}^{(sing)}(L,\Delta=\Delta_c) = c_1 + c_2 L^{(\alpha_s-1)/\nu},
\end{equation}
where $c_1$ and $c_2$ are constants.
Fig. \ref{Cs_crit} is a plot of the excess bond energy, with $J_1/J=0.1$,
 at the bulk critical point. The fit using Eq. (\ref{bondE_form}) 
results in $(\alpha_s-1)/\nu=0.22 \pm 0.03$, corresponding to
\begin{equation}
\alpha_s = 1.30 \pm 0.05.
\end{equation}

\subsection{Magnetization decay close to the surface}

\begin{figure}[ht]
\includegraphics[width=7cm
]{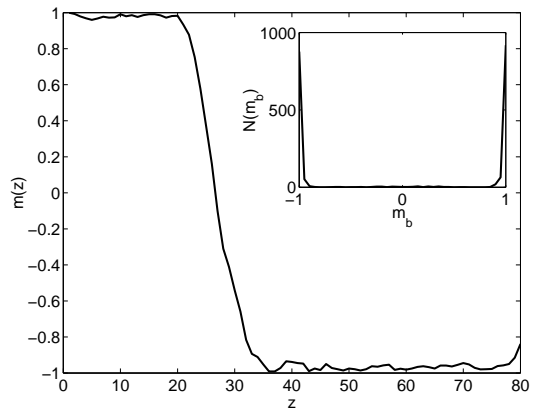}
\caption{Main figure: A typical example of a 
magnetization profile, taken from a single sample,
where due to a strong positive surface 
field $h_1$ at $z=0$ an interface has formed between two regions of 
opposite magnetization.  Inset: Distribution of the bulk magnetization 
$m_b$ with periodic boundary conditions, 2000 samples. 
$\Delta/J=2.27$, system size 40x40x80.}
\label{interface}
\end{figure}

Finally we discuss the behavior of the 
magnetization profiles $m(z)$ (i.e. magnetization as a function of the 
distance $z$ from the surface), in the case the spin orientation at the 
surface layer is fixed. This corresponds to applying a strong surface field $h_1$.
These are of interest as they reflect spin-spin correlations close 
to the surface, as studied in Ref.~\cite{parisisourlas} in the slightly
different context of comparing two replicas with opposite $h_1$. 
For the RFIM close to the infinite 
system bulk critical point, $m(z)$ is affected by the fact that for 
numerically feasible system sizes the bulk magnetization is close to 
unity and decreases very slowly with increasing system size (due to the 
small value of $\beta$) \cite{middleton}. This is demonstrated in 
the inset of Fig. \ref{interface}, where the distribution of bulk 
magnetization $m_b$ at the critical point can be seen to be strongly peaked 
around $m_b = \pm 1$. 

\begin{figure}[ht]
\includegraphics[width=7cm
]{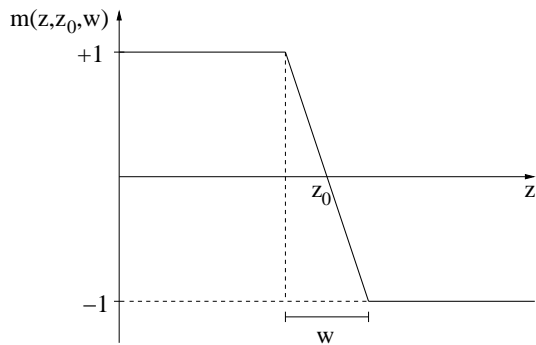}
\caption{A simple model for a single-sample magnetization profile
$m(z,z_0,w)$. The interface is characterized by the parameters
position $z_0$ and width $w$.}
\label{interface_model}
\end{figure}

One can now distinguish three scenarios from sample to sample: 
if $|m_b| \approx 1$ the applied strong surface field 
$h_1$ may have the same or opposite orientation, 
or finally the bulk magnetization $m_b$ may be close to zero.
In the first case, the $h_1$ induced 
spin configuration will be close to the one in the absence of
the field. In the second case, $h_1$ will either force $m_b$ to change 
sign altogether (producing again a flat profile with 
$m(z) \approx \pm 1$) or induce an \emph{interface} between the two 
regions of opposite magnetization, as in Fig. \ref{interface}.
The third one has a small probability, and thus will not
contribute much to the ensemble averaged magnetization profile.
 The \emph{average} magnetization profile $\langle m(z) \rangle$ can then 
(for a finite system, at the infinite system critical point) 
be well approximated by writing
\begin{equation}
\label{mz_ansatz}
\langle m(z) \rangle \approx a + b \langle m_{if}(z) \rangle.
\end{equation}
Here $a$ and $b$ are weight factors, here constant but in
general function(s) of $L$, that tell the relative weight 
of samples where the magnetization changes inside due to the
$h_1$.
\begin{equation}
\label{if_integral}
\langle m_{if}(z) \rangle = \int dw dz_0 P_w(w) P_{z_0}(z_0)m(z,z_0,w)
\end{equation}
is the profile one would obtain by averaging only over ``single
sample'' profiles $m(z,z_0,w)$, corresponding to an interface of width $w$
and position $z_0$ (with probability distributions $P_w$ and $P_{z_0}$,
respectively). A simplified model for $m(z,z_0,w)$ is shown in Fig. 
\ref{interface_model}. 

\begin{figure}[ht]
\includegraphics[width=7cm
]{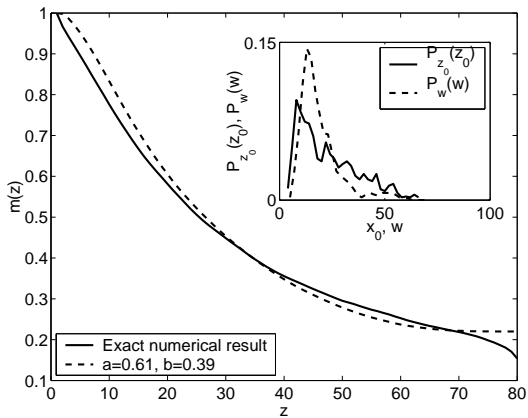}
\caption{Main figure: A comparison between the numerical $\langle m(z) \rangle$
(solid line, averaged over $3000$ samples) and that obtained  by using Eqs. 
(\ref{mz_ansatz}) and (\ref{if_integral}) with $m(z,z_0,w)$ as in Fig. 
\ref{interface_model} (dashed line). Inset: Distributions of the interface 
position $P_{z_0}(z_0)$ (solid line) and width $P_w(w)$ (dashed line) 
obtained from the simulations. $\Delta/J=2.27$, system size 40x40x80.}
\label{profile_comparison}
\end{figure}

From the exact ground state calculations, we identify the profiles 
corresponding to such interface configurations. This is done by demanding 
that such profiles have a region where 
$m(z) < -0.9$ (when $h_1 \gg 0$). The interface width is defined as 
$w = z_2-z_1$, where $z_1$ and $z_2$ are the smallest $z$'s such that 
$m(z_1)<0.9$ and $m(z_2)<-0.9$, respectively. The interface position $z_0$
is then given by $z_0 = (z_1+z_2)/2$. By counting the fraction of 
such profiles, we can estimate $a$ and $b$ in Eq. (\ref{mz_ansatz}). 
These have the approximate values of $0.39$ and $0.61$, respectively 
(for a system of size 40x40x80). By using Eqs. (\ref{mz_ansatz}) and 
(\ref{if_integral}) with $m(z,z_0,w)$ presented in Fig. 
\ref{interface_model}, as well as the distributions $P_w$ and $P_{z_0}$ 
measured from the ground state calculations, one indeed obtains an 
average profile $\langle m(z) \rangle$ that is in reasonable agreement 
with the true one, see Fig. \ref{profile_comparison}. 

The \emph{average} magnetization profile $\langle m(z) \rangle$ decays 
slowly with the distance $z$, not quite reaching zero at the opposite
edge of the system in the case at hand. However, a \emph{typical} 
value of $m(z)$ will be close to $\pm 1$ for all $z$, which persists
for accessible system sizes due again to the small value of $\beta$.
One may thus observe effects reminiscent of violation of 
self-averaging, and this would be true also if one
would measure the averaged difference $\langle |m(z)-m_{GS} (z)|\rangle$
between the field-perturbed and GS 
configurations, and the higher moments thereof. These results
illustrate simply how the quasi-ferromagnetic character of
the 3d RFIM groundstate influences such perturbation studies,
a consequence of the in practice limited system sizes one can
access in simulations.

\section{Conclusions}

In this work we have studied with combinatorial optimization and 
scaling arguments surface criticality in a random magnet, the 3d RFIM.
The surface 
layer magnetization exponent $\beta_1$ is more than an order of magnitude 
larger than the extremely small bulk value
\cite{middleton, rieger, hartmann_m}. 
Experimentalists have reported much larger values for $\beta$ 
\cite{belanger,antifm,kleemann},
which in fact are rather close to our estimate for $\beta_1$.
An intriguing possibility in this respect is the direct observation
of the surface order parameter 
in relaxor ferroelectrics via piezoelectric force microscopy \cite{kleemann2}.

The excess exponents $\alpha_s$ and $\beta_s$,
when inserted into the scaling relations (\ref{alpha_s}) and (\ref{beta_s}),
both yield very small values for the correction term $\nu(\theta-\theta')$,
assuming $\alpha \approx 0$, $\beta \approx 0.02$ and $\nu \approx 1.37$
\cite{middleton}. This suggests that in fact $\theta' = \theta$, and the 
excess exponents are related to bulk exponents by the usual scaling laws 
valid for pure systems, Eq.~(\ref{pure_scaling}). 
The numerically obtained
description of the ordinary surface transition uses the bulk 
correlation length exponent as in pure systems.
All this would
merit further theoretical considerations and could also be checked
in the four-dimensional RFIM \cite{4drfim}, whose phase diagram is
also more complex due to the 3d surfaces which have independently
phase transitions.
The spin-spin correlations close to the surface and the
magnetization profiles in the presence of boundary perturbations
have been studied, similarly to the context of looking for
self-averaging violations \cite{parisisourlas}.
It would be interesting to investigate this aspect in more detail,
but in our numerics the most transparent features are due to the
two-peaked magnetization distribution of the groundstates, without
a perturbing field.

On a final note, the observations here
concerning surface criticality in a disordered magnet - with a 
complicated energy landscape - 
extend directly for instance to spin glasses \cite{spinglasses} and to
a wide class of non-equilibrium systems (see \cite{fran}, also
for experimental suggestions). Two evident possibilities are looking
for the same phenomenology in 3d Ising spin glasses, and in the 3d 
zero-temperature non-equilibrium RFIM. In the former case, the
free surface of a system at $T>0$ is in analogy to the zero temperature
3d RFIM case inherently disordered (the 2d spin glass has a $T=0$ phase 
transition). In the second case, the situation is much more akin to the one
at hand (\cite{fran}) and one should consider as the order parameter
the remanent surface magnetization after a demagnetization procedure.

{\bf Acknowledgments}
A. Hartmann (G\"ottingen),
D. Belanger (Santa Cruz) and W. Kleemann (Duisburg)
are thanked for useful comments, 
and the Center of Excellence program of
the Academy of Finland for financial support.




\end{document}